# Hybrid Optimization of Laser-Driven Fusion Targets and Laser Profiles

Z. Li[1#], Z. Q. Zhao[1#], X. H. Yang[1,4*], G. B. Zhang[1], Y. Y. Ma[2,4], H. Xu[1,4], F. Y. Wu[3,4], F. Q. Shao[1], J. Zhang[3,4]

[1] *College of Science, National University of Defense Technology, Changsha 410073, China*

[2] *College of Advanced Interdisciplinary Studies, National University of Defense Technology, Changsha 410073, China*

[3] *Key Laboratory for Laser Plasmas (Ministry of Education), School of Physics and Astronomy, Shanghai Jiao Tong University, Shanghai, 200240, China*

[4] *Collaborative Innovation Centre of IFSA, Shanghai Jiao Tong University, Shanghai, 200240, China*

*xhyang@nudt.edu.cn

[#]Z. Li and Z. Q. Zhao contributed equally to this work.

**Abstract:**

Quasi-isentropic compression is an effective method to achieve high-density and high-temperature implosion in laser-driven inertial confinement fusion (ICF). However, it requires precise matching between the laser profile and the target structure. Designing the optimal laser profile and the corresponding target for ICF is a challenge due to the large number of parameters involved. In this paper, we present a novel method that combines random walk and Bayesian optimization. The basic sampling data for Bayesian optimization are a series of laser pulse profiles and target structures that can produce relatively high areal densities obtained by the random walk method. This approach reduces the number of samples required for Bayesian optimization and mitigates low efficiency in the latter stages of the random walk method. The method also reduces the randomness in the optimization process and enhances the optimization efficiency. It should have important applications in ICF research.

1. Introduction

Laser-driven inertial confinement fusion (ICF) is a promising approach to achieve controlled nuclear fusion [1, 2, 3, 4]. Recently, significant progress has been achieved, i.e., the National Ignition Facility (NIF) realized the first net energy gain and 3.15 MJ of fusion energy [5]. The main challenge for achieving ignition is to compress the fuel to high enough density and temperature, which are about 50 g/cm$^3$ and 5-12 keV for the hot spot and 300~500 g/cm$^3$ for the cold fuel [6, 7]. Quasi-isentropic compression is the most effective way to achieve high-density compression and reduce the energy requirements.

Nuckolls proposed the theory of using exponential laser profile for quasi-isentropic compression of the target composed of deuterium (D) and tritium (T), and evaluated the potential energy output and gain. When using the ideal isentropic compression laser profile to compress the solid DT spherical target, high gain can be achieved when the laser energy is 1 MJ [1]. Kidder et al. constructed a more complete quasi-isentropic compression model in ICF, demonstrating that isentropic compression can be achieved when the laser is uniformly distributed. They used a novel hollow target design that provides inward acceleration space for thermonuclear fuel, and achieved relatively good compression effects in simulations [8, 9, 10]. Betti and Zhou optimized the target structure to make the ratio of ablator and DT fuel more reasonable. They used a three-stage quasi-isentropic compression laser profile with a flat-top stage, which reduced the peak laser power and improved the compression effect [11]. However, achieving ignition is challenging because various unstable factors can affect the final compression state, such as Rayleigh-Taylor instability (RTI) and laser-plasma instability (LPI) [12]. Radha et al. proposed a laser profile called triple-picket, and finely adjusted the entropy of the target surface through three laser pre-pulses to the growth rate of RTI [13]. Hurricane et al. proposed the theory of a high-foot laser profile, which enhances the ablation rate and entropy of the target to improve the stability of the implosion compression process. [14]. Baker et al. improved the Big-foot laser profile and greatly enhanced the compression effect by increasing the ablation rate and suppressing unstable growth rates [15]. However, the laser profile and target structure have become much more complex in order to suppress instabilities and boost fusion reaction rate. Designing an optimal laser profile and target structure to achieve efficient implosion, i.e., high compression ratio and high temperature, is challenging.

A highly reliable mathematical model can help us to select better parameters from the available ones, thus improving the parameter optimization efficiency. However, given the intricate correlation between the compression implosion effect and the parameters of the ICF laser profile and target structure, it is exceedingly challenging to find parameters that meet the requirements. Machine learning has wide applications in ICF and can solve the parameter adjustment problem [16]. Gopalaswamy et al. applied Bayesian inference to optimize the target and laser profile, resulting in a threefold increase in neutron yield [17]. Peterson et al. applied random forest to train their model and discovered a new type of ICF design capable of achieving high yield even under significant drive asymmetry and non-uniform shell thickness [18]. Wu et al. optimized the laser profile using a genetic algorithm and achieved a 63% areal density increase under similar laser energy [19]. We use the random walk method to design laser profiles and target structures, which can quickly determine the optimal parameters [20]. Machine learning can thus efficiently determine the required laser profile and target parameters for quasi-isentropic compression. However, it relies heavily on a significant amount of experimental or simulation data, and most ICF research groups struggle to accumulate such a vast amount of data.

We present a hybrid optimization method for designing laser profiles and target structures for ICF by combining random walk and Bayesian optimization [21, 22]. The method aims to optimize the laser profile and target structure under a given laser energy to achieve a higher areal density, neutron yield, ion temperature, and other performance metrics. The method is introduced firstly and is benchmarked with some previously published results and is proved to be a very efficient target designing method. Then it has been applied to design the laser profile and target structure for the Double-Cone Ignition (DCI) experiment on the SGII upgrade laser facility [23].

## 2. The hybrid optimization method

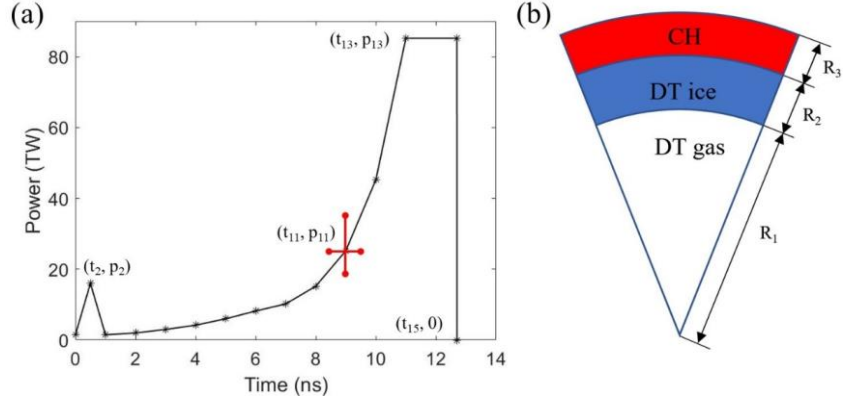

Fig. 1 The laser profile (a) and the three-layer target structure (b). The thick red line indicates the time and power ranges of the 11th point in the laser profile, the others are similar to it.

It can achieve efficient implosion compression for ICF by using the ideal isentropic compression laser profile derived by Kidder [24]. The laser profile to be optimized in this work are shown in Figure 1(a), which is described by 15 points. The duration of $t_1 \sim t_3$ is the pre-pulse to improve the hydrodynamic instability [25, 26], $(t_3, p_3) \sim (t_8, p_8)$ is the duration of the foot for laser profile, $(t_8, p_8) \sim (t_{13}, p_{13})$ corresponds to the duration of the ramp, and $(t_{13}, p_{13}) \sim (t_{15}, t_{15})$ is the duration of the flat-top [27]. The flat-top duration reduces the peak laser power, mitigates the laser-plasma instability growth and enhances the compression [28]. Both the pre-pulse power $p_2$ and the laser peak power $p_{13}$ are strictly controlled in the optimization process. Both $t_1$ and $p_{15}$ are set to zero and $p_{14}$ is set to $p_{13}$. The laser profile is designed under a given laser energy, so $t_{14}$ depends on the previous variables rather than being an independent parameter and $t_{15} = t_{14} + 0.001$ ns. Therefore, 25 independent parameters are used to describe the laser profile.

Figure 1b shows the typical target structure for the direct-driven ICF. The target consists of three layers from outside to inside: CH ablator, DT ice, and DT gas [29]. The dense CH material ($\rho$ =1.1 g/cm$^2$) effectively compresses the less dense DT ice ($\rho$ =0.25 g/cm$^2$) and DT gas ($\rho$ =0.0003 g/cm$^2$) inside. The thick DT gas layer provides a long acceleration distance for the outer DT ice layer. Three independent parameters are used to describe the target structure: the radius of DT gas ($R_1$), and the thicknesses of the DT ice ($R_2$) and CH ($R_3$) layers.

Similar laser intensity ($I_L$), adiabatic factor ($\alpha$), and implosion velocity ($V_i$) can be obtained by scaling the existing laser profile and target structure using the hydro-equivalent theory [30]. For a given laser energy ($E_L$), the target inner radius (R), the ablation layer thickness ($\Delta$), the laser peak power ($P_L$), and laser main pulse duration ($t_L$) are related as follows: $R \propto E_L^{1/3}$, $\Delta \propto E_L^{1/3}$, $P_L \propto E_L^{2/3}$ and $t_L \propto E_L^{1/3}$. The initial laser profile and target structure can be set by using the scaling, and applied as the starting point for the optimization.

## 2.1 Random walk optimization

We optimize the laser profile and target structure by combining the radiation hydrodynamic code MULTI-IFE [31] and the random-walk method [20] firstly. MULTI-IFE is a 1D spherical symmetry code, which incorporates the essential physics for the implosion and thermonuclear ignition, e.g., two-temperature hydrodynamics of ions and electrons, three-dimensional laser light ray-tracing, thermal diffusion, multigroup radiation transport, nuclear reaction of DT, and transport of alpha-particles. The opacity is calculated by the SNOP code, which is based primarily on the average atomic model [32]. The equation of state (EOS) is given by the FEOS code [32], which includes the free electron energy, bounding electron energy, and ion energy.

To narrow down the selection range of the new parameters, based on the parameters of the previous generation, the variation range of each generation is set. $X_{i,n}$ (i = 1, 2, ...; n = 1, 2, …) denotes the parameters, where i is the index of the time, power, radius, or thickness, and n is the generation number. The range of $X_{i,n}$ for the $n^{th}$ generation is [$(X_{i-1, n-1}+X_{i, n-1})/2$, $(X_{i, n-1}+X_{i+1, n-1})/2$], and the variation range of each parameter changes dynamically with the optimization process. However, this new set of parameters cannot be used directly for calculation. Let $u_{i,n} = X_{i,n} - X_{i, n-1}$ (i = 1, 2, ..., n = 1, 2, ...), perform a normalization process, and then multiply by the step size to obtain $X_{i,n} = X_{i, n-1} + u_{i,n} * S$ [20].

After being modified, the parameters are loaded into the MULTI-IFE code for computation. The step size S halves if the current iteration produces better results than the previous one; otherwise, it remains unchanged. The next generation parameters are generated using the aforementioned method based on the new generation parameters, and this process is repeated until the step size S becomes very small, then the optimization is ended.

The optimization method described above has several advantages: 1) As the step size gradually decreases, it can effectively obtain a series of input parameters, and the corresponding results are mostly close to the optimal value. This ensures that a large amount of data can be collected to achieve quasi-isentropic compression, and drastic changes in the parameters that may fall outside the effective range can be avoided simultaneously. 2) The parameter interval is updated dynamically with each generation, avoiding the manual delineation of the parameter range. However, when the result approaches to the optimal one, the increase of step size becomes tiny, preventing the optimal parameters being identified. In the cases, combining the random walk optimization with Bayesian optimization can solve the aforementioned problems and improve the efficiency of the optimization process [33].

## 2.2 Bayesian optimization

Bayesian optimization is a very effective global optimization algorithm that can quickly and efficiently find the relatively optimal input parameters. The objective function is often a black-box function that its exact expression form or derivatives are unknown. Gaussian Process is the foundation of Bayesian optimization [33]. A Gaussian Process is uniquely defined by its mean function and covariance function [34].

The Gaussian process assumes a normal distribution for the given parameters and models the unknown function with uncertainty as a surrogate function. By applying the Bayesian formula P(A|B) = P(B|A)*P(A)/P(B), we can obtain the posterior distribution from the prior distribution. The P(A|B), denoted as posterior distribution, represents the distribution that requires optimization. Meanwhile, P(B|A) represents the objective function, P(A) is the prior distribution of the unknown function, and P(B) represents the evidence or marginally possible data [35]. After several iterations, the Bayesian optimizer based on Gaussian process will approach the function with a certain confidence level, and gradually enhance the confidence level and converge to the optimal solution.

The optimization of laser profile and target structure is considered as an example. We model the relationship between the laser profile, target structure, and areal density as y = f(X), where X represents the relevant parameters

$$X = \begin{bmatrix} X_1 \\ ... \\ X_n \end{bmatrix} \quad (1)$$

y = ρR represents the maximum areal density achieved at the end of laser irradiation. This allows us to transform the problem of finding the optimal areal density into an optimization problem within a boundary domain A:

$$X^* = \arg max_{X \epsilon A} f(X) \quad (2)$$

Bayesian optimization requires initial sampling to obtain a dataset $X_0$. Given a mean function and a kernel function, Gaussian process regression can then provide a succinct representation of the data set X~ f(X) by using the mean vector and the covariance matrix. Note that the choice of kernel function can significantly affect the estimation of the unknown function.

The radial basis function (RBF) kernel is used here, and its smooth property that enables it to capture the nonlinearity and is suitable for optimizing the laser profile and target structure [36]. The RBF is defined as:

$$k(X_1, X_2) = \sigma^2 \exp\left(-\frac{\|X_1 - X_2\|^2}{2l^2}\right) \quad (3)$$

where $X_1$ and $X_2$ are two different sets of parameters, σ is the variance and l is the scale length of the RBF kernel, which are typically set to 1 [36].

An initial function $f_1$ is obtained using the mean and kernel functions, but it may not be optimal due to the limited data $X_0$ and high uncertainty [35]. To improve the estimate, additional observed data points are introduced along with the previous ones. This reduces the uncertainty and provides better design variables. Various methods can be used to choose the next data point that balances exploration and exploitation, which can be determined using an acquisition function α(X) in Gaussian process regression [37]. In this work, the Confidence Bound criteria method is adopted to avoid the local optima and find a new optimal solution based on the current estimate [35].

The random walk optimization method can generate a large number of parameters within the desired ranges. By using these parameters as the initial sampled data for Bayesian optimization, the number of samples required for Bayesian optimization can be significantly decreased. Moreover, Bayesian optimization overcomes the limitation of the random walk optimization method, i.e., it becomes harder to optimize the parameters as the step size decreases during the optimization process. The hybrid optimization method combines both of the advantages of the algorithms and is more suitable for optimizing the laser profile and target structure for ICF. Figure 2 shows the flow chart of hybrid optimization.

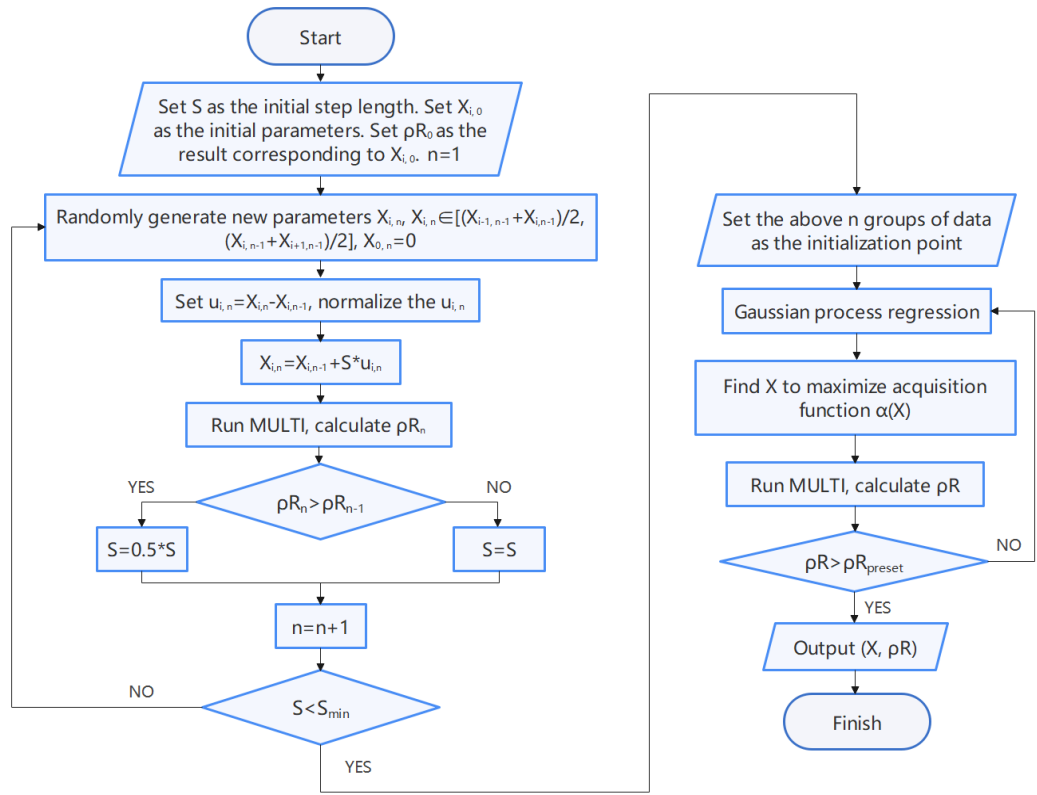

Fig. 2 Flow chart of hybrid random walk and Bayesian optimization method.

## 3. Results and discussion

### 3.1 Target optimization with different laser energy

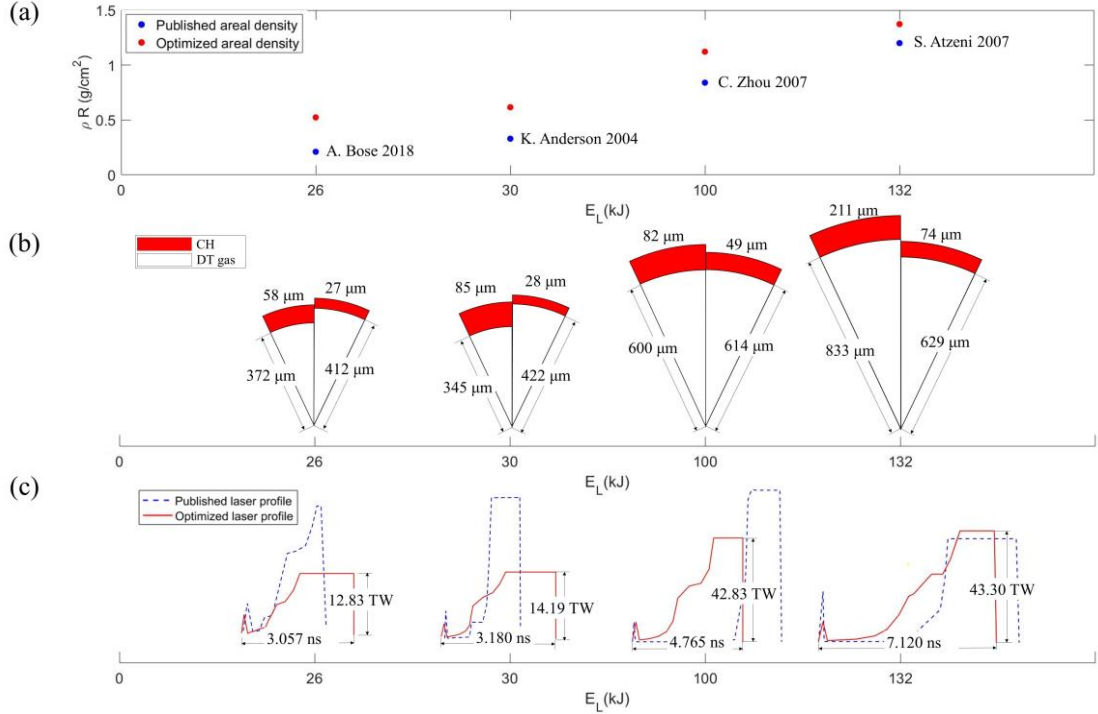

Fig. 3 The areal densities given by the hybrid optimization and the references are marked with red circles and blue squares (a), respectively. The target structure from the references (left) and the hybrid optimization method (right) (b). The laser profile from the references (blue dotted line) and the hybrid optimization method (red solid line) (c).

Figure 3 shows the design and optimization of four sets of laser profiles and target structures with different laser energies for ICF. All targets are double-layered structure, consisting of DT gas and CH ablation layer. The choice of laser energy follows the work by Betti et al. [25, 39, 40, 41], and we try to match their initial parameters of the laser profile and target structure as closely as possible. Figure 3a shows the comparison between our optimized results and their published data. The hybrid optimization method increases the areal density by 0.2-0.3 g/cm$^2$ with a smoother laser profile.

Figure 4 shows that the computation required for hybrid optimization increases with the laser energy. It is attributed to the larger range of parameters that needs to be chosen for the laser power, the pulse width, and the target size. More calculations are needed to find the optimal parameters. The areal density also increases with the 1/3 power of the laser energy. In addition, the proportion of random walk optimization and Bayesian optimization in the hybrid optimization varies with different laser energies. For fusion schemes with lower laser energy, the computational cost of the random walk method is significantly higher than that of Bayesian optimization during the hybrid optimization process. The reason for this is that the random walk method entails a step size halving process. It takes several generations of calculation for the step size to become small enough, and only then can the optimization process end.

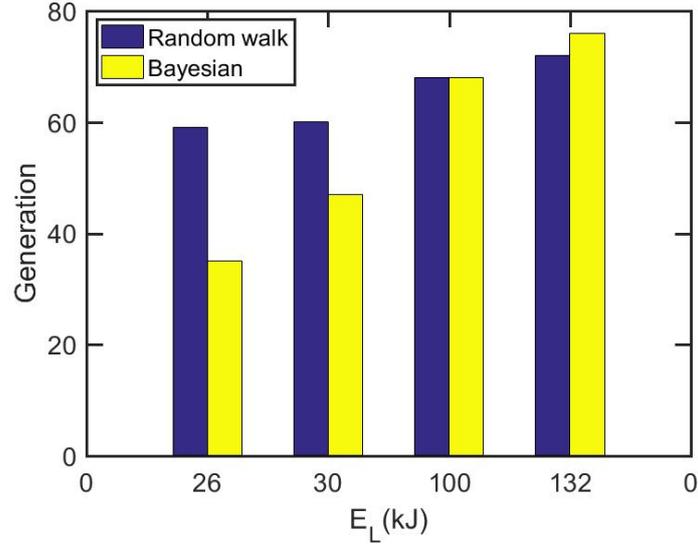

Fig. 4 The iteration generations for the random walk optimization and the Bayesian optimization, respectively, corresponding to the four hybrid optimizations shown in Figure 3.

As shown in Figure 3(b), compared with the parameters obtained from the theory of hydro-equivalent ignition, in schemes with lower laser energy, the optimized targets have a larger outer radius, while in schemes with higher laser energy, the optimized targets have a smaller outer radius. The optimized ablation layer thicknesses also decrease, leading to an increase of the implosion velocity. For the lower energy cases, the optimized laser pulse profiles have a lower peak power and a longer pulse width, as shown in Figure 3 (c). In the optimization process, the laser intensity on the target surface is maintained under 800 TW/cm$^2$ to suppress the laser plasma instability (LPI) [42], so that the peak power of the laser profile decreases for all the schemes. As the peak power decreases, the pulse duration increases, resulting in a longer acceleration for the target. The increase of the target inner radius also leads to a longer acceleration distance for the target, resulting in an enhanced compression and higher areal density.

As the laser energy increases, the pulse width of the optimized laser pulse decreases in the cases with higher laser energies. This is also related to the smaller size of the optimized target. 132 kJ and 336 kJ belong to the same order of magnitude, so that the peak power and pulse width of the laser profile in the optimized 132 kJ scheme are closer to the original design. Therefore, the theory of hydro-equivalent ignition is more suitable for ICF schemes with similar orders of magnitude of laser energy and can serve as a reference for future ICF designs.

**3.2 Optimization of 80 kJ laser energy for DCI**

The hybrid optimization method is an efficient tool to design complex target structures and high-energy laser fusion. The target structure and the laser pulse profile for a 60 kJ laser energy for DCI [23] are designed and optimized by applying it. The initial CH ablation layer have a thickness of 86 μm, and the initial DT ice layer have a thickness of 97 μm. The low-density DT gas have a radius of 862 μm. The initial laser profile has a peak intensity of 105.5 TW, and a duration of around 10 ns. The pre-pulse has a peak intensity of around 25 TW, and a duration of approximately 0.5 ns. These are suitable initial parameters that could be further optimized using the method described in Section 2. The goal is to keep the laser intensity within a reasonable range and maximize the peak areal density.

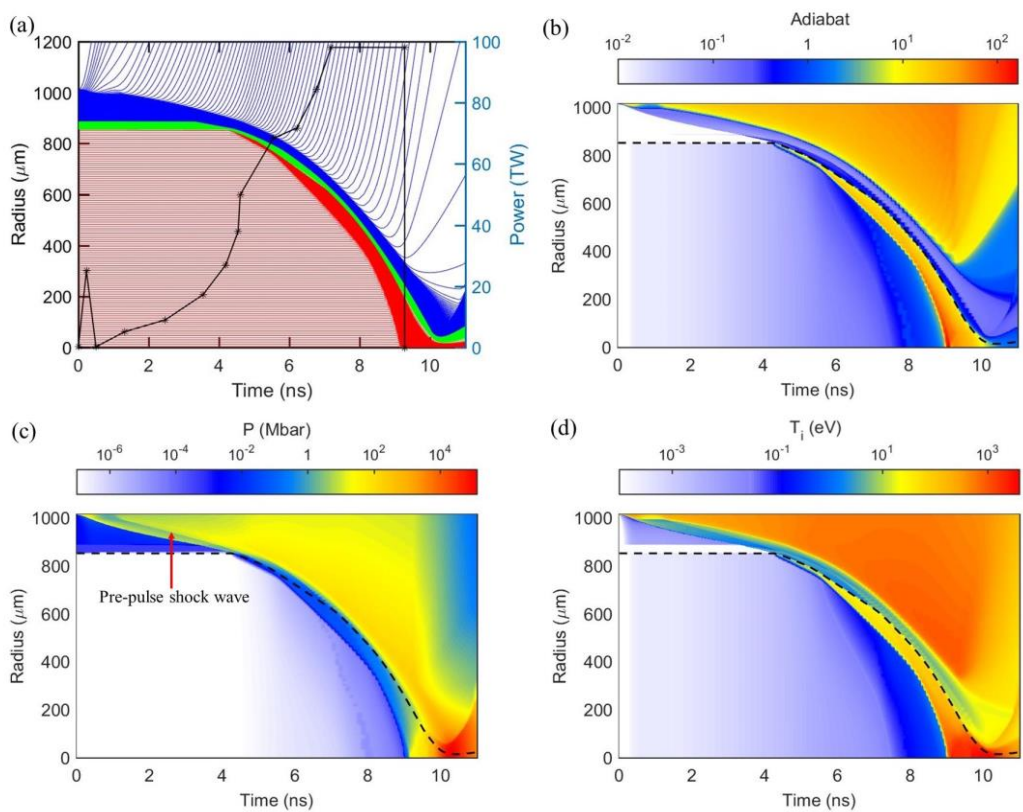

Fig. 5 The implosion diagram (a), the laser profile is also presented. The evolution of the adiabat with the time (b). The evolution of pressure with the time, in which the shock propagation is shown clearly (c). The evolution of the ion temperature with the time (d). The black dashed line is the interface of the DT ice and the DT gas.

The optimal values for each input parameter are determined. The optimal inner radius of the target is 855 μm, and the optimal thicknesses of the DT ice layer and CH layer are 33 μm and 125 μm, respectively. The laser peak power reaches 85.20 TW with a pulse width of 9.27 ns. The peak implosion velocity is about 236 km/s. Figure 5a shows that the optimized laser pulse has a shorter pulse duration than that of the theoretical quasi-isentropic compression laser profile. This is because the ramp duration is relatively long, and a considerable portion of the laser energy is allocated during this duration. The target begins to be accelerated and compressed inward during half of the ramp duration, and the in-flight aspect ratio IFAR = $R_1/(R_2 + R_3)$ at the initial acceleration stage is the largest, reaching 19.03. It is lower than the optimal range of 35-40 [13], the outer layer of the target is also the thinnest at this time. To prevent the shock wave penetrating into the DT gas and increase of the entropy, we need to slow down the rising speed of laser power at this time. Then, the laser power increases again until it reaches its peak power and enters the flat-top duration. During the stagnation phase, the DT reaches its peak density, which are 253.6 g/cm$^3$ and 76.55 g/cm$^3$ for DT ice and DT gas, respectively. When the DT begins to expand outward, the outer ablation layer still has a tendency to compress inward due to the inertia. At this time, the thickness of the ablation layer reaches its minimum value and the peak density reaches 131.5 g/cm$^3$. Figure 5b shows that the pre-pulse increases the entropy of the DT ice layer. At 5.57 ns, the entropy of DT gas near the cold fuel increases to some extent in response to the shock wave, but the DT gas further inside is still in a low-entropy state. The internal DT gas remains low-entropy during the pre-pulse, which enhances ablation rate and reduces the RTI [25]. Figure 5c shows that the target undergoes a shock wave in the outer layer during the initial compression stage because of the pre-pulse. Figure 5d shows the evolution of the ion temperature during the implosion. The temperature of the hot spot reaches its peak value of 3.62 keV. After the laser power turns off, the temperature of the outer surface of the target begins to gradually decrease. However, since the target is still being compressed towards the center by its inertia, the temperature of the internal DT gas is still increasing.

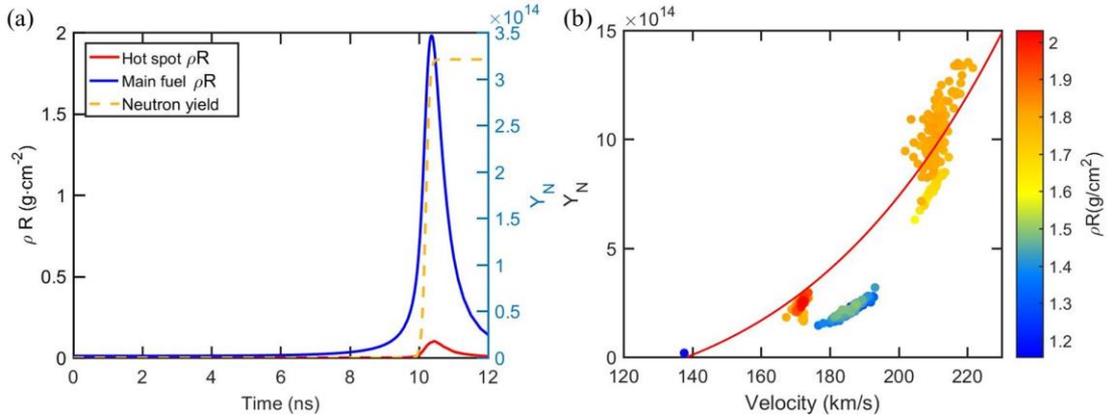

Fig. 6 The evolution of the areal density and the neutron yield with respect to the time (a). The evolution of the areal density and the neutron yield with the peak implosion velocity during the hybrid optimizations (b).

As shown in Figure 6a, the total areal density reaches 2.03 g/cm$^2$. According to the hydro-equivalent theory [30], the total areal densities of the target shell and hot spot are given by the hydrodynamic relationship in inertial confinement fusion,

$$\rho R_{\text{shell}} (\text{g}/\text{cm}^2) \approx \frac{1.2}{\alpha^{0.54}} \left[ \frac{E_{\text{L}}(\text{kJ})}{100} \right]^{0.33} \left[ \frac{V_{\text{i}}(\text{cm}/\text{s})}{3 \times 10^7} \right]^{0.06} \quad (4)$$

$$\rho R_{\text{hs}} (\text{g}/\text{cm}^2) \approx \frac{0.31}{\alpha^{0.55}} \left[ \frac{E_{\text{L}}(\text{kJ})}{100} \right]^{0.27} \left[ \frac{V_{\text{i}}(\text{cm}/\text{s})}{3 \times 10^7} \right]^{0.62} \quad (5)$$

where α is the in-flight adiabat, $E_{\text{L}}$ is the laser energy and $V_{\text{i}}$ is the peak implosion velocity. To account for the pre-pulse, the in-flight adiabat is adjusted to 2 in the initial compression [23]. By substituting the new implosion velocity and adiabat into Eqs. (4) and (5), the maximum total areal density corresponding to the implosion velocity is 1.61 g/cm$^2$, which is lower than the optimal result. That is, the hybrid optimization enables better design of laser profiles and target structures. Figure 6b shows the relationship between the neutron yield and the peak implosion velocity in the hybrid optimization process, which is basically consistent with $Y_N \propto V_i^{4.2}$ [17]. The cases with lower areal densities are owing to the mismatch of the laser pulse profile and the target structure. Therefore, their neutron yield and peak implosion velocity do not follow this relationship. This further verifies the effectiveness of hybrid optimization.

Figure 7 shows that the areal density increases significantly in the first stage of the random walk optimization method. However, as the number of calculations increases and the step size decreases gradually, the increase of the areal density slows down and the optimization efficiency decreases. For the Bayesian optimization method, we first define the variation range for each parameter as indicated by the red line shown in Figure 1. After 200 Bayesian optimization, the areal density increases steadily with the iteration number, but only reaches 1.41 g/cm$^2$ at the end. In contrast, with the hybrid optimization that applying the data of the 98$^{th}$ generation of random walk optimization as the basis of Bayesian optimization, the maximum areal density is obtained after only 96 generations of calculations.

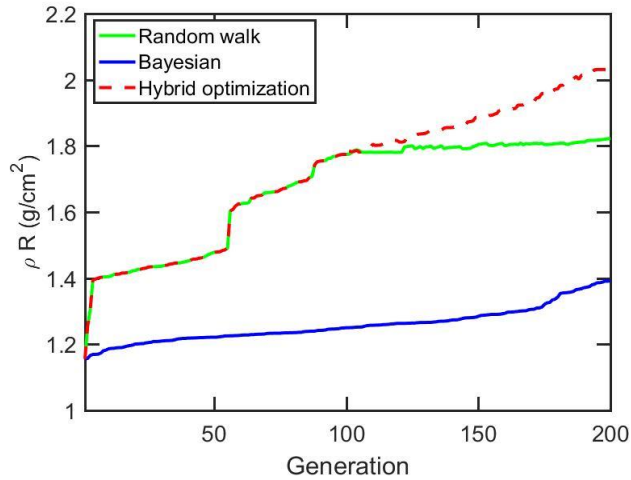

Fig. 7 The iteration generation for the three optimization methods.

## 4. Conclusion

In summary, we have proposed a new method for optimizing the structure of fusion targets and the laser pulse profile under a given laser energy. By combining random walk and Bayesian optimization methods, it can efficiently design the target structure and laser pulse profile for different laser energies, aiming to achieve high areal densities or other desired parameters. The results show that this method can improve the areal density by 0.2-0.3 g/cm$^2$ comparing to that in the previous works, and exceed the prediction of the hydrodynamic theory. This method is suitable for the complex laser profile designing that suppresses the Rayleigh-Taylor instability (RTI) [13], thus is practical for engineering applications, such as the DCI experiments performed on the SGII upgrade laser facility [23].

**Acknowledgments**

This work was supported by the National Natural Science Foundation of China (Grant Nos. 12175309, 11975308, 12005297, and 12275356), the Strategic Priority Research Program of Chinese Academy of Science (Grant No. XDA25050200 and XDA25010100), X.H.Y. also acknowledges the Fund for NUDT Young Innovator Awards (No. 20180104).

**Data Availability**

The data that support the findings of this study are available from the corresponding author upon reasonable request.